\documentclass[epj]{svjour}
\usepackage{graphics}
\begin{document}
\title{The high-energy galactic tau neutrino flux
and its atmospheric background}
\author{Husain Athar\inst{1} \and  Kingman Cheung\inst{2} \and Guey-Lin Lin\inst{3}
\and Jie-Jun Tseng\inst{4}
}                     
\institute{Physics Division, National Center for Theoretical
Sciences, Hsinchu 300, Taiwan \and Department of Physics, National
Tsing-Hua University, Hsinchu 300, Taiwan \and Institute of
Physics, National Chiao-Tung University, Hsinchu 300, Taiwan \and
Institute of Physics, Academia Sinica, Taipei 115, Taiwan }
\date{Received: date / Revised version: date}
%
\abstract{ We compare the tau neutrino flux arising from the
galaxy and the earth atmosphere for $10^{3} \leq E/\mbox{GeV} \leq
10^{11}$. The intrinsic and oscillated tau neutrino fluxes from
both sources are considered.  We find that, for $E\geq 10^3$ GeV, the
oscillated $\nu_{\tau}$ flux along the galactic plane dominates
over the maximal intrinsic atmospheric $\nu_{\tau}$ flux, i.e.,
the flux along the horizontal direction. We also briefly comment
on the prospects for observing these high-energy tau neutrinos.
\PACS{
      {95.85.Ry}{Neutrino, muon, pion, and other
      elementary particles; cosmic rays}   \and
      {13.85.Tp}{Cosmic-ray interactions}
     } 
} 
\maketitle
\section{Introduction}
\label{intro} The Milky way is one of the nearby astrophysical
sources producing high energy neutrinos, besides the familiar
earth atmosphere \cite{Athar:2003gw}. Measurements of galactic
neutrino and photon fluxes could provide information about the
distribution of matter and cosmic rays in the galaxy. Furthermore,
the above flux is also a background for the search of more distant
high energy neutrino sources such as the AGNs and the GRBs. In
this talk, we shall focus on the flux of $\nu_{\tau}$. It is clear
that the observation of astrophysical $\nu_{\tau}$, with a flux
comparable to the flux of $\nu_{e}$ and $\nu_{\mu}$, directly
confirms the neutrino oscillations. In order to observe galactic
$\nu_{\tau}$ flux, it is essential to study the atmospheric
background. We shall focus on the energy range $E_{\nu}\geq 10^3$
GeV.

We first discuss the distinction between intrinsic and oscillated
neutrino fluxes arising from the Milky way and the earth
atmosphere. We then present results for galactic and atmospheric
tau neutrino fluxes. Finally we comment on the prospects for
observing galactic tau neutrinos.
\section{Intrinsic and oscillated neutrino fluxes}
\label{sec:1}
It is well known that the relative flavor ratio for
astrophysical neutrinos at the source is approximately
$\phi_{\nu_e}^0:\phi_{\nu_{\mu}}^0 :\phi_{\nu_{\tau}}^0=1:2:0$.
The commonly accepted astrophysical processes for producing electron and muon
neutrinos are $(\gamma,p)+p\to \pi^{\pm}+X$ where pion further
decays into electron and muon neutrinos with the ratio
$\phi_{\nu_e}^0:\phi_{\nu_{\mu}}^0=1:2$. On the other hand, the production
mechanism for the tau neutrino is $(\gamma,p)+p\to D_s+X$ where $D_s$ further
decays into tau neutrinos. The production cross section for $D_s$ meson is much smaller
than that for $\pi^{\pm}$ for center-of-mass energy $\sqrt{s}\sim $ (1-10) GeV.
 Hence $\phi_{\nu_{\tau}}^0$ is suppressed compared to
$\phi^{0}_{\nu_e}$ and $\phi^{0}_{\nu_{\mu}}$.

Although $\nu_{\tau}$ flux is rather suppressed at the source, it
is not negligible at the detector. As neutrinos propagate to the earth, the
oscillation effect takes place and the relative neutrino flavor ratio
changes as a result.  Let us denote the neutrino flux reaching the
earth as $\phi_{\nu_{\alpha}}$. Then \cite{Athar:2000yw}
\begin{equation}
\phi_{\nu_{\alpha}}=\sum_{\beta}P_{\alpha\beta}
\phi_{\nu_{\beta}}^0,
\end{equation}
where $P_{\alpha\beta}$ is a function of neutrino mixing matrix
$U_{\alpha i}$ which connects the neutrino mass eigenstate to the
flavor eigenstate. The $\alpha, \beta $ run over $e, \mu$ and $\tau$.
 Assuming a bi-maximal mixing for $U_{\alpha
i}$, one obtains for vanishing $\delta $ and $\theta_{13}$ \cite{Athar:2000ak}
\begin{equation}
P_{\alpha\beta}=\left(
\begin{array}{ccc}
 1/2 & 1/4 & 1/4 \\
 1/4 & 3/8 & 3/8 \\
 1/4 & 3/8 & 3/8
\end{array}
\right).
\end{equation}
Such a probability matrix implies $\phi_{\nu_e}:
\phi_{\nu_{\mu}}:\phi_{\nu_{\tau}}=1:1:1$. This flavor ratio is
applicable to galactic neutrinos since these neutrinos propagate
through a distance much greater than the neutrino oscillation
length. On the other hand, this ratio is not applicable to the
atmospheric neutrinos for $E_{\nu}\geq 10^3$ GeV considered here.
At this energy, the oscillation length for the atmospheric
neutrino is of the order $10^{10}$ cm, which is much greater than
even the earth diameter. Hence the atmospheric $\nu_{\tau}$ flux
for $E_{\nu}\geq 10^3$ GeV must be dominantly an {\em intrinsic}
one.
\section{The galactic and atmospheric tau neutrino
fluxes}
\label{sec:2}
The total galactic tau neutrino flux consists of
intrinsic and oscillated components. Both components follow from
the collisions of primary cosmic-ray proton with interstellar
medium proton.  The density of interstellar medium proton is taken
to be $n_p=1\, {\rm cm}^{-3}$ along the galactic plane, while the
primary cosmic ray spectrum,
 $\phi_{p}(E_{p})$, is taken to be \cite{JACEE}
%
%
\begin{equation}
\label{proton}
 \phi_{p}(E_{p})= \left \{
           \begin{array}{ll}
          1.7 \, (E_{p}/\mbox{GeV})^{-2.7} & {\rm for}\; E_{p}<E_{0}, \\
          174 \, (E_{p}/\mbox{GeV})^{-3}   & {\rm for}\; E_{p}\geq
          E_{0},
        \end{array}
      \right.
\end{equation}
where $E_{0}=5\cdot 10^{6}$ GeV and $\phi_{p}(E_{p})$ is in units
of cm$^{-2}$ s$^{-1}$ sr$^{-1}$ GeV$^{-1}$. We assume directional
isotropy in $\phi_{p}(E_{p})$ for the above energy range. A more
recent measurement of cosmic-ray flux spectrum between $2\cdot
10^{5}$ GeV and $10^{6}$ GeV agrees with the $\phi_{p}(E_{p})$
given by Eq. (\ref{proton}) within a factor of $\sim $ 2 in this
energy range \cite{Amenomori:2000br}. The intrinsic galactic tau
neutrinos are produced by $p+p\to D_s+X$, with the $D_s$ meson
decays into a $\tau$ lepton and a $\nu_{\tau}$, while the $\tau$
lepton further decays into the second $\nu_{\tau}$
with other particles. We calculate the $D_s$ production cross
section by the following two approaches: (i) the perturbative QCD
(PQCD) and (ii) the quark-gluon string model (QGSM)
\cite{Kaidalov:xg}.  In the PQCD approach, we use the CTEQ5 parton
distribution functions \cite{Lai:1999wy} and apply a $K$ factor, $K=2$, to account
for the NLO corrections \cite{Nason:1989zy}.
%
%
%
%
%
With the $D_s$
production cross section determined, one can calculate the
$\nu_{\tau}$ flux using
\begin{equation}
\label{AA}
\phi_{\nu_{\tau}}^0=
 \int_{E}^{\infty}
 \mbox{d}E_p \; \phi_{p}(E_{p}) \, f(E_{p}) \, \frac{1}
 {\sigma_{pp}(E_p)} \;
 \frac{\mbox{d}\sigma_{pp \to \nu_{\tau}+Y}}{\mbox{d}E},
\end{equation}
where $f(E_{p})=R/\lambda_{pp}(E_{p})$ with $\lambda_{pp}(E_{p})$
the $pp$ interaction length and $R$ a representative distance in
the galaxy along the galactic plane. We take $R$ to be $\sim 10$
kpc, where 1 pc $\simeq 3\cdot 10^{18}$ cm.
 We have focused on the intrinsic tau neutrino flux along the
galactic plane just to obtain the maximal expected tau neutrino
flux. The matter density decreases exponentially in the direction
orthogonal to the galactic plane, therefore the amount of
intrinsic tau neutrino flux decreases by approximately two orders
of magnitude for the energy range of our interest.
\begin{figure}
\resizebox{0.4\textwidth}{!}{%
\includegraphics{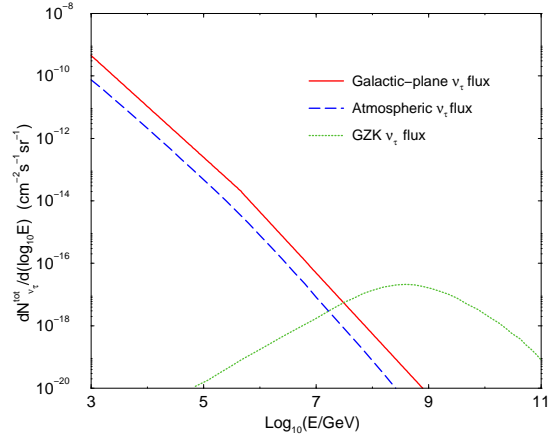}
}
\caption{Galactic-plane, horizontal atmospheric and GZK tau
neutrino fluxes under the assumption of neutrino flavor
oscillations.}
\label{fig:1}       
\end{figure}
Another component of galactic tau neutrinos comes from the
oscillation of galactic muon neutrinos. The flux of the latter can
also be calculated by Eq.~(\ref{AA}) with $\nu_{\tau}$ replaced by
$\nu_{\mu}$ \cite{Stecker:1978ah}. In this case,  $\pi^{\pm}$ and
$K^{\pm}$ are the dominant intermediate states that decay into
muon neutrinos. Since $\phi_{\nu_{\mu}}^0$ is few orders of
magnitude greater than $\phi_{\nu_{\tau}}^0$ while
$\phi_{\nu_e}^0$ is approximately one half of
$\phi_{\nu_{\mu}}^0$, we recover the flavor ratio
$\phi_{\nu_e}^0:\phi_{\nu_{\mu}}^0 :\phi_{\nu_{\tau}}^0=1:2:0$ for
the galactic neutrinos at the source. From the previous section,
we have $\phi_{\nu_{\tau}}=\phi_{\nu_{\mu}}^0/2$.  The flux  $\phi_{\nu_{\tau}}$ can be
parameterized as
%
%
\begin{equation}
\label{parameterization}
 E\phi_{\nu_{\tau}}(E)= \left \{
           \begin{array}{ll}
          1.5\cdot 10^{-5} \, (E/\mbox{GeV})^{-2.63} & {\rm for}\; E<E_{1}, \\
          9.5\cdot 10^{-4} \, (E/\mbox{GeV})^{-2.95}   & {\rm for}\; E\geq E_{1},
        \end{array}
      \right.
\end{equation}
where $E_{1}=4.7\cdot 10^{5}$ GeV.
 The $E\phi_{_{\nu_{\tau}}}(E)$ is in units of cm$^{-2}$ s$^{-1}$ sr$^{-1}$.

 Having discussed the calculation of galactic tau neutrinos, we now
turn to the atmospheric tau neutrinos. As said before, for
$E_{\nu}\geq 10^3$ GeV, atmospheric $\nu_{\tau}$ flux has only the
intrinsic component. We have used the nonperturbative QCD approach
mentioned earlier to model the production of $D_{s}$ mesons in the
$pA$ interactions. We have used the $\phi_{p}(E_{p})$ given by Eq.
(\ref{proton}) and the $Z$-moment description for the calculation
of intrinsic tau neutrino flux \cite{Gaisser:vg}.
We obtain the atmospheric tau neutrino flux by solving a set of cascade equations \cite{ACLT,ext}

The results for galactic and
atmospheric tau neutrino fluxes, along with the GZK
\cite{Greisen:1966jv} oscillated tau neutrino flux
\cite{Engel:2001hd}, are presented Fig. 1, where we have used
 the notation $\phi_{\nu_{\tau}}
 \equiv {\mbox d}N_{\nu_{\tau}}/{\mbox d}(\log_{10}E)$ . From the figure, we
note that the galactic plane oscillated $\nu_{\tau}$ flux {\em
dominates} over the intrinsic atmospheric $\nu_{\tau}$ flux for
$E\leq 5\cdot 10^{7}$ GeV, whereas the GZK oscillated tau neutrino
flux dominates for $E\geq 5\cdot 10^{7}$ GeV. Quantitatively, the
atmospheric $\nu_{\tau}$ flux in the horizontal direction is $\sim $ 5
times smaller than the galactic-plane $\nu_{\tau}$ flux.
Furthermore, the downward atmospheric $\nu_{\tau}$ flux is factor of
 $\sim 8$ smaller than its horizontal counterpart. Let us recall
here that particle physics aspects of intrinsic tau neutrino flux
calculation presented here are empirically supported only up to
 $\sqrt{s}\sim $ TeV, which corresponds to $E_{p}\sim 10^{6}$ GeV.
\begin{figure}
\resizebox{0.4\textwidth}{!}{%
\includegraphics{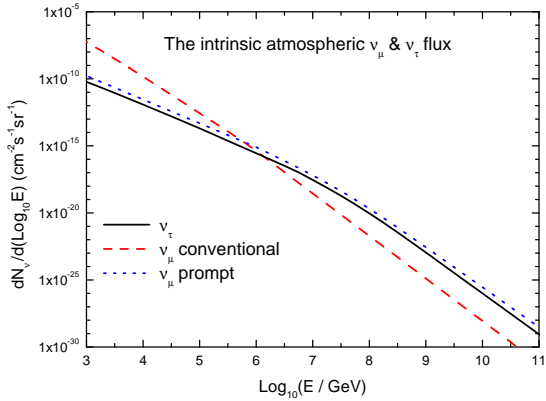}
}
\caption{The comparison of downward atmospheric $\nu_{\mu}$ and
$\nu_{\tau}$ fluxes}
\label{fig:2}       
\end{figure}

Before we move on, it is instructive to compare the atmospheric
$\nu_{\tau}$ and $\nu_{\mu}$ fluxes. We plot these two fluxes for
$E_{\nu}\geq 10^3$ GeV in Fig. 2. The atmospheric $\nu_{\mu}$ flux is taken from
Ref.~\cite{Gondolo:1995fq}.
The atmospheric $\nu_{\mu}$ flux has two components: conventional
and prompt components. The former component is due to decays of
$\pi^{\pm}$ and $K^{\pm}$, while the latter component is due to
decays of charm hadrons. Although produced more copiously than
charm hadrons, $\pi$ or $K$ meson does not decay efficiently in
the air at sufficiently high energy. This effect disfavors the
resulting neutrino flux. Indeed, in Fig. 2, the two components of
$\nu_{\mu}$ flux cross roughly at $E_{\nu}=10^6$ GeV
\cite{Gondolo:1995fq}. We also observe that the atmospheric
$\nu_{\mu}$ and $\nu_{\tau}$ fluxes are comparable for $E_{\nu}>
10^6$ GeV, since both fluxes are due to charm production in this
energy range.

\section{The prospects of observations}
\label{sec:3}

 For downward going or near horizontal high-energy
tau neutrinos, the deep inelastic neutrino nucleon scattering,
occurring near or inside the detector, produces two showers
\cite{Learned:1994wg}. The first shower is due to a
charged current neutrino nucleon deep inelastic scattering,
whereas the second shower is due to the (hadronic) decay of the
associated $\tau$ lepton produced in the first shower. It might be
possible for the proposed large neutrino telescopes such as the
IceCube to constrain the two showers simultaneously for
$10^{6}\leq E/\mbox{GeV} \leq 10^{7}$, depending on the achievable
shower separation capabilities \cite{Athar:2000rx}.
 Here, the two showers develop mainly in ice. With such a detection
strategy, we estimate the event rate for observing the galactic
tau neutrinos. In the above energy range for the tau neutrinos,
the event rate in 1 km$^3$ water/ice  Cherenkov detector is
rather small, about $\sim 5\cdot 10^{-3}\, {\rm yr}^{-1}\, {\rm
sr}^{-1}$. Such a low event rate implies that one can only search
for galactic tau neutrinos in the lower energy. Alternatively, one
 should consider other detection strategies. Since
Fig.~\ref{fig:1} shows that galactic tau neutrino flux dominates
over its atmospheric counterpart for $E_{\nu}\geq 10^3$ GeV, it is
desirable to develop strategies for identifying tau neutrinos at
TeV energies or even lower because of relatively large absolute
$\nu_{\tau}$ flux. The present AMANDA search for all flavor
neutrino-induced cascades is not tight enough to constrain/observe
 our predicted $\phi_{\nu_{\tau}}$ \cite{Ahrens:2002wz}.

We point out that the persistent dominance of galactic tau
neutrino flux over its atmospheric background is a unique
phenomenon among all neutrino flavors. Such a dominance does not
occur for $\nu_e$ and $\nu_{\mu}$. For example, the result in Sec.
2 tells us that $\phi_{\nu_{\mu}}=\phi_{\nu_{\tau}}$ for the
galactic neutrinos. On the other hand, the atmospheric $\nu_{\mu}$
flux is much greater than the atmospheric $\nu_{\tau}$ flux for
$E_{\nu}< 10^5$ GeV, as can be seen from Fig. 2. Hence the
galactic $\nu_{\mu}$ flux no longer dominates over the atmospheric
$\nu_{\mu}$ flux for $E_{\nu}< 10^5$ GeV.

In conclusion, we have presented our calculations of both the
galactic and atmospheric tau neutrino fluxes for $E_{\nu}\geq
10^3$ GeV. The former flux is shown to dominate over the latter.
Such a dominance is unique to
 {\em tau neutrinos}. The event rate for galactic tau neutrinos by
observing the double showers (with $10^{6}\leq E_{\nu}/\mbox{GeV}
\leq 10^{7}$) is rather suppressed. Therefore, to observe the
galactic tau neutrinos, it is desirable to develop techniques for
identifying tau neutrinos at lower energies.

H.A. and K.C. are supported in part by the Physics Division of
National Center for Theoretical Sciences under a grant from the
National Science Council of Taiwan. G.L.L. and J.J.T. are
supported by the National Science Council of Taiwan under the
grant numbers NSC91-2112-M-009-019 and NSC91-2112-M-001-024.
%
%

\end{document}